\documentclass[letter]{ptptex}

\usepackage{amsmath}

\title{
 On the Equation of Motion for a Fast Moving Small Object in the
 Strong Field Point Particle Limit%
}

\author{
Takashi \textsc{Fukumoto},$^1$ 
 Toshifumi \textsc{Futamase}$^1$ and  Yousuke \textsc{Itoh}$^2$%
}

\inst{
$^1$Astronomical Institute, Graduate School of Science, \\
Tohoku University, Sendai 980-8578, Japan\\
$^2$Physics Department, University of Wisconsin Milwaukee \\
P. O. Box 413, Milwaukee WI 53201, USA}

\recdate{April 10, 2006}

\abst{
We present a straightforward, divergence-free derivation of the equation of motion for a small but finite object in an arbitrary background using the strong field point particle limit. The resulting equation is that of a generalized  geodesic for a non-rotating spherical object. This is consisitent with the results of previous studies.}

\begin{document}
\maketitle


\newcommand{\edth}{\raisebox{-0.2mm}{$\;\: \tilde{}$} \hspace{-2.2mm}
\raisebox{0.0mm}{$\partial$}}

\vspace{0.5cm}


Equations of motion in general relativity have attracted much interest recently because of the belief that it may be possible to 
 use gravitational waves as  an astronomical tool in the near future. 
In fact, several gravitational wave detectors are now operating around the world, 
~\cite{LIGO, GEO, TAMA} and there are plans to build the next generation
detectors. Gravitational waves may open a new window to the universe,
just as radio waves and X-rays did. Gravitational waves should allow us to study entirely new regimes in the universe, 
such as the surfaces of black holes and the very beginning of the universe,
that no other type of radiation can reveal, 
and they will probably allow us to find new astronomical objects that we have not imagined to
exist.   
However, in order to realize this, not only technology but also theory must be developed further. 
Theoretically, it is necessary to have accurate predictions of the nature
 of the waves emitted by various astrophysical sources. 
For this purpose, we need to have a good understanding 
of equations of motion with radiation reaction in a wide variety of situations. 
In particular, at present we have only a very poor understanding of the equation of motion for a fast 
moving source, such as small but extended
objects (e.g., neutron stars or black holes), in an arbitrary external field. 
In such a situation, the object does not necessarily move along a geodesic of the external field, 
because the self-field of the object and the gravitational waves generated by the orbital 
motion are not negligible. 

In this paper, we study the equation of motion describing such a situation. 
It should be mentioned that Mino {\it et al.} and others have derived an equation of motion 
for a point particle with mass $m$ that is represented by a delta function source 
in an arbitrary background, and the equation is interpreted as a 
geodesic equation on the geometry determined by the external field and the so-called tail part 
of the self field of the particle to first order in $m$~\cite{Mino,QW, DW}.  
Furthermore, Mino {\it et al.} used another approach, the matched asymptotic 
expansion, to obtain an equation of motion without employing the concept of a point particle and thus 
avoided divergence in their derivation.  

Our approach is different from the previously used approaches for the reason that we take the extended nature of the object 
into account and thus avoid using a singular source. This is done by useing the point 
particle limit developed by one of the present authers\cite{Futamase}. 
We believe that our approach simplifies the proof given by Mino {\it et al.} (employing the matched asymptotic expansion) that the
Mino-Sasaki-Tanaka equation of motion is applicable to a nonsingular
source.

Let us start by explaining the situation we consider and introduce the strong field point particle 
limit,~\cite{Futamase} which plays an essential role in our formulation.  
A small compact object 
with mass $m$ moves with arbitrary speed near a massive body with mass $M$. 
We assume that the object is stationary, except for higher-order tidal effects, so that 
we can safely ignore 
emission of gravitational waves from the object itself. However, of course, we cannot ignore the 
gravitational waves emitted by the orbital motion of the object. 
We denote the world line of the center of mass of the object by $z(\tau)$ and define 
the body zone of the object as follows. We imagine a spherical region around $z(\tau)$ whose radius scales as $\epsilon$. 
Also, we stipulate that the linear dimension of the object scale as $\epsilon^2$, so that the boundary of the body 
zone is located at the far zone of the object. Thus, we are able to carry out a multipole expansion of the 
field generated by the object at the surface of the body zone. 
We also implicitly assume that the mass of the object scales as $\epsilon^2$, so that the compactness 
of the object remains constant in the point particle limit
$\epsilon \rightarrow 0$. 
This is why we call this limit the strong field point particle limit.
Then we calculate the metric perturbation induced by the small object in this limit.  
The smallness parameter $\epsilon^2$ has the dimension of length, and this length characterizes the smallness of the object. 
One may regard it as the ratio of the physical scale of the object
and the characteristic scale of the background curvature. 
We assume that the
background metric  $g_{\mu\nu}$ satisfies the Einstein equations in vacuum. Therefore, the Ricci
tensor of the background vanishes. Since we have assumed that mass scale of the
small object is much smaller than the scale of gravitational field of the
background geometry, we approximate the metric perturbation by the
linear perturbation of the small particle, $h_{\mu\nu}$.

We employ the harmonic gauge, in which we have 
\begin{equation}
\bar{h}^{\mu\nu}_{\,\,\,\,\,\,;\nu}=0,
\end{equation}
where the semicolon represents the covariant derivative with respect to the
background metric, and the trace-reversed variable is defined as usual: 
\begin{equation}
\bar{h}_{\mu\nu}=h_{\mu\nu}-\frac{1}{2}g_{\mu\nu}g^{\rho\lambda}h_{\rho\lambda}.
\end{equation}
Then the linearized Einstein equations take the form
\begin{equation}
 -\frac{1}{2}\bar{h}^{\mu\nu;\xi}_{\quad;\xi}\left(x\right)-R^{\mu\,\,\,\nu}_{\,\,\,\xi\,\,\,\rho}\left(x\right)\bar{h}^{\xi\rho}\left(x\right)=8\pi
 T^{\mu\nu}\left(x\right).
\end{equation}
This can be solved formally as 
\begin{equation}
\bar{h}^{\mu\nu}\left(x\right)=8\pi\int d^4 y\sqrt{-g}\,G^{\mu\nu}_{\quad\alpha\beta}\left(x,y\right)T^{\alpha\beta}\left(y\right), 
\end{equation} 
where we have used the retarded tensor Green function defined by
\begin{equation}
G^{\mu\nu\alpha\beta}\left(x,y\right)=\frac{1}{4\pi}\theta\left(\Sigma\left(x\right),y\right)
\left[u^{\mu\nu\alpha\beta}\left(x,y\right)\delta\left(\sigma\left(x,y\right)\right)
+v^{\mu\nu\alpha\beta}\left(x,y\right)\theta\left(-\sigma\left(x,y\right)\right)\right].
\end{equation}
For a general tensor Green function and the definitions of 
$\sigma\left(x,y\right)$, $\bar{g}^{\mu\alpha}\left(x,y\right)$,
$u^{\mu\nu\alpha\beta}\left(x,y\right)$, $v^{\mu\nu\alpha\beta}\left(x,y\right)$,
and $\Sigma\left(x\right)$, please refer to Refs. ~\cite{DB}and  ~\cite{Mino}\,.

Now we take the point particle limit.  
In this limit, the above metric perturbation contains terms with different $\epsilon$ dependences. 
This simplifies the calculation of the equation of motion.
For example,  terms with negative powers of $\epsilon$ appear from the delta function part of 
the Green function, 
\begin{equation}
_{s}\bar{h}^{\mu\nu}\left(x\right)=2\int d^4y \sqrt{-g}\,u^{\mu\nu}_{\quad\alpha\beta}
\left(x,y\right)\delta\left(\sigma\left(x,y\right)\right)T^{\alpha\beta}\left(y\right).
\end{equation}

As explained below, we only need the field on the boundary of the body zone, which is the far zone of 
the body itself, 
and thus we may make use of the multipole expansion for the field.  
For this purpose, we choose our coordinates as
$$y^{\alpha}=z^{\alpha}\left(\tau\right)+\delta y^{\alpha},$$
$$z^{0}\left(\tau\right)=\tau,\quad\delta y^{0}=0,$$
where $z^{\alpha}\left(\tau\right)$ is the world line of the center of the object. 
The center is assumed always to be 
inside the body in the point particle limit, and thus there is no ambiguity regarding the choice of the center. 
In these coordinates, the volume element satisfies $d^4 y=d\tau d^3 \delta y$. Thus we have 
\begin{equation}
_{s}\bar{h}^{\mu\nu}\left(x\right)=2\int d^3\delta y
 \sqrt{-g}\,u^{\mu\nu}_{\quad\alpha\beta}\left(x,z\left(\tau_y\right)+\delta
 y\right)\frac{1}{\dot{\sigma}|_{\tau=\tau_y}}T^{\alpha\beta}\left(y\right),
\end{equation}
where $\tau_y$ is the retarded time of each point $y$.
Then the multipole expansion is obtained by expanding the above expression at the retarded time of 
the center of the object $\tau_z$, 
defined by $\sigma\left(x,z\left(\tau_z\right)\right)=0$.
This can be done easily by using 
the condition $\sigma\left(x,z\left(\tau_y\right)+\delta y\right)=0$. Then the diference between 
$\tau_z$ and $\tau_y$ is given by
\begin{equation}
\delta \tau=-\frac{\sigma_{;\alpha}\left(x,z\left(\tau_z\right)\right)\delta
 y^{\alpha}}{\dot{\sigma}\left(x,z\left(\tau_z\right)\right)}+O\left(\delta y^2\right).
\end{equation}
Using this $\delta \tau$, we can expand  $u^{\mu\nu}_{\quad\alpha\beta}\left(x,z\left(\tau_y\right)+\delta y\right)$ 
about $\tau_z$. In principle, we can calculate arbitrary higher multipole moments in this way. 
Here we only calculate the first two 
terms, namely the monopole and spin terms. 
Then we only need the following expressions:
\begin{equation}
\begin{split}
&u^{\mu\nu}_{\quad\alpha\beta}\left(x,z\left(\tau_y\right)+\delta y\right)\\
&=u^{\mu\nu}_{\quad\alpha\beta}\left(x,z\left(\tau_z\right)\right)
+\left(u^{\mu\nu}_{\quad\alpha\beta;\gamma}\left(x,z\left(\tau_z\right)\right)
-\frac{\sigma_{\gamma}\left(x,z\left(\tau_z\right)\right)}{\dot{\sigma}\left(x,z\left(\tau_z\right)\right)}\frac{du^{\mu\nu}_{\quad\alpha\beta}}{d\tau}\right)\delta y^{\gamma},
\end{split}
\end{equation}
and 
\begin{equation}
\begin{split}
&\dot{\sigma}\left(x,z\left(\tau_y\right)+\delta y\right)\\
&=\dot{\sigma}\left(x,z\left(\tau_z\right)\right)
\left[1+\left(\frac{\dot{\sigma}_{;\gamma}\left(x,z\left(\tau_z\right)\right)}
{\dot{\sigma}\left(x,z\left(\tau_z\right)\right)}-\frac{\ddot{\sigma}\left(x,z\left(\tau_z\right)\right)
\sigma_{;\gamma}\left(x,z\left(\tau_z\right)\right)}{\dot{\sigma}^2\left(x,z\left(\tau_z\right)\right)}\right)
\delta y^{\gamma}\right].
\end{split}
\end{equation}

By defining the mass and the spin as
\begin{equation}
m\dot{z}^{\alpha}\left(\tau_z\right)\dot{z}^{\beta}\left(\tau_z\right)=
\int d^3 \delta y \sqrt{-g} \,T^{\alpha\beta}\left(y\right),
\end{equation}
\begin{equation}
mS^{\gamma(\alpha}\left(\tau_z\right)\dot{z}^{\beta)}\left(\tau_z\right)
=\int d^3 \delta y \sqrt{-g} \,\delta y^{\gamma}\,T^{\alpha\beta}\left(y\right),
\end{equation}
(see Ref. ~\cite{Dix} for the definition of the spin tensor), we finally obtain the following expression for  $_{s}\bar{h}^{\mu\nu}$:
\begin{equation}
\begin{split}
_{s}\bar{h}^{\mu\nu}\left(x\right)&=\frac{2m}{\dot{\sigma}\left(x,z\left(\tau_z\right)\right)}u^{\mu\nu}_{\quad\alpha\beta}\left(x,z\left(\tau_z\right)\right)\dot{z}^{\alpha}\left(\tau_z\right)\dot{z}^{\beta}\left(\tau_z\right)\\
&+\frac{2m\ddot{\sigma}\left(x,z\left(\tau_z\right)\right)\sigma_{;\gamma}\left(x,z\left(\tau_z\right)\right)}{\dot{\sigma}^3\left(x,z\left(\tau_z\right)\right)}u^{\mu\nu}_{\quad\alpha\beta}\left(x,z\left(\tau_z\right)\right)S^{\gamma\alpha}\left(\tau_z\right)\dot{z}^{\beta}\left(\tau_z\right)\\
&+\frac{2m}{\dot{\sigma}\left(x,z\left(\tau_z\right)\right)}u^{\mu\nu}_{\quad\alpha\beta;\gamma}\left(x,z\left(\tau_z\right)\right)S^{\gamma\alpha}\left(\tau_z\right)\dot{z}^{\beta}\left(\tau_z\right)\\
&-\frac{2m}{\dot{\sigma}^2\left(x,z\left(\tau_z\right)\right)}\frac{d}{d\tau}\left(u^{\mu\nu}_{\quad\alpha\beta}\left(x,z\left(\tau_z\right)\right)\sigma_{;\gamma}\left(x,z\left(\tau_z\right)\right)\right)S^{\gamma\alpha}\left(\tau_z\right)\dot{z}^{\beta}\left(\tau_z\right).
\end{split}
\end{equation}

Now we derive the equation of motion using this expression. 
First, we define the $\epsilon$ dependent 4-momentum 
of the object as the volume integral of the effective stress energy tensor $\Theta^{\mu\nu}$ over 
the body zone $B(\tau)$:
\begin{equation}
P^\mu(\tau)=- \int_{B(\tau)} \Theta^{\mu\nu} d\Sigma_\nu.  
\end{equation}   
Here we choose the Landau-Lifshitz form for the stress energy tensor, 
\begin{equation}
\Theta^{\mu\nu}=(-g)(T^{\mu\nu}+t^{\mu\nu}_{LL}), 
\end{equation}
where $T^{\mu\nu}$ is the stress-energy tensor of matter and $t^{\mu\nu}_{LL}$ the Landau-Lifshitz(LL) psuedo-tensor, 
whose explicit expression can be found in their textbook~\cite{LL}.
Because the effective stress energy tensor satisfies the conservation 
law $\Theta^{\mu\nu}_{\, \, \, \, ,\nu}=0$, 
the change of the 4-momentum defined above can be expressed as a surface 
integral over the boundary 
of the body zone $\partial B$:
\begin{equation}
\frac{dP^\mu} {d\tau} 
 =- \oint_{\partial B}  \epsilon^2 n_\nu (1 + \epsilon a \cdot n )
 (-g)t^{\mu\nu}_{LL} d\Omega. 
\end{equation} 
Here, $n^{\mu}$ is the unit normal vector to the surface and $a^{\mu}=D u^{\mu}/d\tau$ the 4-acceleration.  
The equation of motion is obtained by taking the point particle 
limit, $\epsilon \rightarrow 0$, in this expression. 
Thus, we need to calculate $(-g)t^{\mu\nu}_{LL}$ on the body zone boundary (field points), 
on which the multipole expansion 
of the self field and the Taylor expansion of the nonsingular part of the field are known, 
and to retain only terms of order $\epsilon^{-2}$ in the expression.
All the other terms vanish in the limit or 
the angular integration. 

For simplicity, we demonstrate the derivation of the equation of motion in the case of a non-rotating, spherical object.
Recalling that the LL tensor is bilinear in the Christoffel symbol and 
that the Christoffel symbol is a derivative of the metric tensor, it is found that the only remaining 
terms come from the combination of the 0-th order of 
the smooth part of the metric and the part of the self field proportional to $\epsilon^{-1}$, which takes the following form: 
\begin{equation}
^{\left(-1\right)}_{\quad \,\,s}\bar{h}^{\mu\nu}(x) 
= {{2m}\over {\dot \sigma(x, z(\tau))}}u^{\mu\nu}_{\quad\alpha\beta}(x, z(\tau)) 
\dot{z}^{\alpha}(\tau)\dot{z}^{\beta}(\tau)|_{\tau=\tau_{z}}.
\end{equation}  
The remaining part of the self field is the so-called tail part, which is regular 
in the limiting process. 
The field point $(x, \tau_x)$ is now on the body zone boundary,   
which is defined by $\sqrt{2\sigma(x, z(\tau_x))}=\epsilon$ and  
\begin{equation}
[\sigma_{;\alpha}(x, z(\tau)){\dot z}^\alpha(\tau)]_{\tau=\tau_x}=0. 
\end{equation}
We have assumed a stationary spherical source, and in this case, $m$ is the Arnowitt-Deser-Misner (ADM) mass 
that the compact object would have if it were isolated,
\begin{equation}
m=\lim_{\epsilon \rightarrow 0} P^{\tau}.
\end{equation}

Using the above expressions in the LL tensor,  we have the following result in the point particle limit:
\begin{equation}
\frac{dP^\mu} {d\tau}=- m \Gamma^{\mu}_{\alpha\beta} (g_s)u^\alpha u^\beta 
- m \frac{1}{2} \Gamma^{\alpha}_{\alpha\beta} (g_s)u^{\beta}u^{\mu}.
\end{equation} 
Here the Christoffel symbol  is evaluated using the 0-th order of the smooth 
part of the metric $g_s$.  
Using the fact that the ADM mass is related to the 4-momentum as
\begin{equation}
P^\mu = \sqrt{-g_s}\, m u^\mu 
\end{equation}
(which is suggested by the higher-order post Newton approximation~\cite{Itoh}), we finally have
\begin{equation}
\frac{du^\mu} {d\tau}=- \Gamma^{\mu}_{\alpha\beta} (g_s)u^{\alpha}u^{\beta}, 
\end{equation} 
which is the geodesic equation on the geometry determined by the smooth part of the metric around the 
compact object.

In fact, the spin effect on the equation of motion can be derived in a
similar way, and the standard result is obtained~\cite{TH}.

We have proved that a small compact object moves on the geodesic determined by 
the smooth part of the geometry around an object. The equation describing this motion is the so-called MiSaTaQuWa equation. 
The smooth part contains gravitational waves emitted by 
the orbital motion, and therefore the equation contains the damping force due to the 
radiation reaction. 
It is necessary to explicitly calculate the smooth part of the metric generated by the small object around the 
orbit. This is the heart of the problem, and it is not attempted in this paper. 
However, we point out that our method avoids use of a 
singular source, employing only the retarded Green function, 
by making use of the point particle limit, and all the quantities are evaluated at the surface of the body zone boundary. Thus we only need 
the dependence of the distance from the center of the object, namely, the $\epsilon$ dependence of 
the field. For these reasons we are able to avoid using any divergent quantities in any part of 
our calculation. One of the difficulties encountered when using a singular source is the ambiguity of the separation between the divergent 
and convergent parts at the singularity.  With our method, it might be possible to avoid this kind of ambiguity and to obtain a unique equation 
for the fast motion with a radiation reaction. 
We would like to see this if this is indeed the case in a future work.

\bigskip
\begin{center}
{\bf Acknowledgements}
\end{center}
\bigskip

T. F and Y. I. would like to thank Prof. B. Schutz for 
useful discussions.
Y. I. would like to thank Prof. M. Shibata for his hospitality 
during Y. I.'s stay at Tokyo University. 
This work was supported in part by a Grant-in-Aid for Scientific Research (No.15540248) of 
Ministry of Education, Culture, Sport, Science and Technology in Japan, 
as well as by the COE program at Tohoku University.

\end{document}